\documentclass[pra,final,twocolumn,superscriptaddress]{revtex4-1}
\usepackage{graphicx,psfrag}
\usepackage{color}

\begin{document}

\title{Fermi-liquid Landau parameters for a nondegenerate band:\\
        Spin and charge instabilities in the extended Hubbard model}

\author{     Gr\'egoire Lhoutellier}
\affiliation{Laboratoire CRISMAT, UMR 6508 CNRS-ENSICAEN, Caen, France}

\author{     Raymond Fr\'esard}
\affiliation{Laboratoire CRISMAT, UMR 6508 CNRS-ENSICAEN, Caen, France}

\author{     Andrzej M. Ole\'s}
\affiliation{Marian Smoluchowski Institute of Physics, Jagiellonian
             University, prof. S. \L{}ojasiewicza 11, PL-30348 Krak\'ow, Poland }
\affiliation{Max-Planck-Institut f\"ur Festk\"orperforschung,
              Heisenbergstrasse 1, D-70569 Stuttgart, Germany}

\date{\today}

\begin{abstract}
We investigate the Landau parameters for the instabilities in spin
and charge channels in the nondegenerate extended Hubbard model with
intersite Coulomb and exchange interactions. To this aim we use the
spin rotationally invariant slave boson approach and we determine the
necessary inverse propagator matrix. The analytically derived spin
Landau parameter $F_0^a$ for the half filled
band uncovers the intrinsic instability of the nondegenerate Hubbard
model towards ferromagnetism --- negative intersite exchange
interaction triggers a ferromagnetic instability
at half filling before the metal-insulator transition, indicated by the
divergence of the magnetic susceptibility at $F_0^a=-1$. This result
is general and the instability occurs in the
strongly correlated metallic regime for any lattice, in three or two
dimensions. Next as an illustrative example we
present numerical results obtained for the cubic lattice
with nearest neighbor exchange $J$ and Coulomb $V$ elements and
arbitrary electron density. One finds that the range of small doping
near half filling is the most unstable one towards spin polarization,
but only in the case of ferromagnetic intersite exchange $J<0$. Charge
Landau parameter $F_0^s$ is lowered near half filling by increasing
$U$ when the intersite Coulomb interaction $V$ is attractive, but in
contrast to $F_0^a$ at $J<0$ it requires an attraction beyond a
critical value $V_c$ to generate the divergence of the charge
susceptibility at $F_0^s=-1$ in the metallic phase.
This instability was found for a broad range of electronic filling
away from half filling for moderate attraction.
\end{abstract}


\maketitle

\section{Introduction}
\label{sec:int}

The Hubbard model has been initially
introduced, \textit{inter alia}, to describe metallic magnetism
\cite{Hub63,Kan63}. This is frequently forgotten in recent research
which focuses on other instabilities which are hidden and not so easy
to investigate, as the instabilities of the Fermi surface or the onset
of superconductivity \cite{Walter}. The ferromagnetic (FM) state is
easily derived from the Stoner criterion and occurs at sufficiently
large local Coulomb interaction $U$. The Stoner instability of a
nonmagnetic state is determined using the Hartree-Fock approximation
and is thus predicted for any lattice and filling of a nondegenerate
band. However, it was soon realized that it does not imply
ferromagnetism as electron correlations strongly renormalize the Stoner
parameter which follows from the Hartree-Fock approach and remove the
FM instability in most cases. The metallic state of strongly correlated
electrons in the nondegenerate Hubbard model is a correlated Fermi
liquid which is characterized by the Fermi liquid parameters. In the
frequently considered model of a rectangular density of states, the FM
instability is absent. This conclusion follows also by considering the
Fermi-liquid antisymmetric parameter $F_0^a$, which lower bound was
found to be $-3/4$ \cite{Vol84}. Indeed, ferromagnetism in a
nondegenerate band is rare and only very few systems exist in nature
which could be classified as itinerant ferromagnets with a single band,
such as a weak itinerant ferromagnet ZrZn$_2$ \cite{Yel05}.

After several decades of research on the Hubbard model, it became clear
that the mechanism of itinerant ferromagnetism is subtle and other
effects are needed to stabilize it. One of them is high degeneracy of
the ground state \cite{Tas92} realized when the band is flat
\cite{Tas94,Tas98,Der14}. Ferromagnetism in realistic flat-band systems
has been proposed for a triangular lattice of Na$_x$CoO$_2$ \cite{Mer06}
and for the $p$ orbitals in honeycomb lattices with ultracold atoms
\cite{Zha10}. In a two-dimensional (2D) Hubbard model on a square
lattice FM instability occurs at the van Hove singularity of the
density of states \cite{Fle97,Hlu97,Zim98}. It becomes relevant away
from half filling when the band is flat due to the next neighbor
hopping $t'$. This follows from a more general approach which predicts
the FM instability by investigating the divergence of the uniform
magnetic susceptibility.

Other factors which stabilize itinerant ferromagnetism in real
materials go beyond the Hubbard model itself --- these are:
(i)~direct intersite exchange coupling \cite{Hir89,Str94,Bar04}, or
(ii)~band degeneracy with active Hund's exchange coupling
\cite{Sto90,Lamb}.
The second mechanism is better known --- it is responsible for
ferromagnetism in transition metals Fe, Co and Ni which have rather
large magnetic moments and FM states occur in a degenerate $3d$ band
where Hund's exchange plays a prominent role \cite{Vol97}.
Also here the Stoner parameter is strongly renormalized by the
electron correlation effects \cite{Sto90}, but the FM instability
survives and is much easier to obtain than for the nondegenerate $s$
band \cite{Fle97}. FM states have been investigated for realistic
transition metals using different methods including more recent studies
with the Gutzwiller wave function \cite{Bun98}. This subject is
interdisciplinary and itinerant ferromagnetism plays also a role in
a ferromagnet/insulator/superconductor ballistic junction \cite{Ann09},
while the essential role played in the onset of ferromagnetism by the
orbital degeneracy was also demonstrated recently for the multiband
Hubbard models on square and cubic lattices \cite{Li14} relevant both
for $p$-orbital bands with ultracold fermions in optical lattices,
and for electronic $3d$-orbital bands in transition-metal oxides
\cite{Czy94}.

Here we focus on the extended nondegenerate Hubbard model which
describes the simplest correlated metals and investigate the influence
of intersite Coulomb and exchange interactions on the charge and
magnetic instabilities. Thereby we follow the route initiated by
Vollhardt \cite{Vol84} who investigated Landau Fermi liquid parameters
$F_0^a$ and $F_0^s$ for $^3$He. Further studies gave the Fermi liquid
interaction and the quasiparticle scattering amplitude on the Fermi
surface in $^3$He \cite{Pfi86}. The perturbative method predicts the
change of sign of the parameter $F_0^a$ in the 2D square
lattice for small $U\simeq 2t$ \cite{Fus00}, where $t$ is the hopping
parameter. The calculation within the slave boson method in the doped
2D Hubbard model \cite{Kag96} does not confirm this result and gives
a negative $F_0^a$ in the entire range of $U$ and its saturation with
increasing $U$ towards a negative value
but larger than $-1$ and nearly independent of the doping.
This behavior confirms the earlier result that the nondegenerate
Hubbard model does not exhibit a FM instability even when $U$ is very
large \cite{Vol84}. We remark that saturated ferromagnetism near half
filling and at $U\to\infty$ which stems from the Nagaoka state is
nevertheless not excluded, as shown by several studies
\cite{Tas98,vdL} including a slave boson approach \cite{Doll2}.
This subject goes however beyond the scope of the present study.

Since Mott insulating ground states arise at large $U$ and at half
filling, we perform our investigations in a framework which is able to
capture interaction effects beyond the physics of Slater determinants.
It is an extension of the Kotliar and Ruckenstein slave boson
representation that reproduced the Gutzwiller approximation on the
saddle-point level \cite{Kot86}. It entails the interaction driven
Brinkman-Rice metal-to-insulator transition \cite{Bri70}. A whole range
of valuable results have been obtained with Kotliar and Ruckenstein
\cite{Kot86} and related \cite{Li89,FW} slave boson representations
which motivate the present study.
In particular they have been used to describe
antiferromagnetic \cite{Lil90}, ferromagnetic \cite{Doll2}, spiral
\cite{Fre91,Arr91,Fre92} and striped \cite{SeiSi,Fle01,Sei02,Rac06}
phases. Furthermore, the competition between
the latter two has been addressed as well \cite{RaEPL}. The influence
of the lattice geometry on the metal-to-insulator transition was
discussed, too \cite{Kot00}. For instance, a very good agreement with
Quantum Monte Carlo simulations on the location of the
metal-to-insulator transition for the honeycomb lattice has been
demonstrated \cite{Doll3}. Finally, further motivation comes from
the strongly inhomogeneous polaronic states that have been found in correlated
heterostructures using the Hubbard model extended with intersite
Coulomb interactions \cite{Pav06}.

Furthermore, comparison of ground state energies to existing numerical
solutions have been carried out for the square lattice, too. For
instance, for $U=4t$ it could be shown that the slave boson ground
state energy is larger than its counterpart by less than 3\%
\cite{Fre91}. For larger values of $U$, it has been obtained that the
slave boson ground state energy exceeds the exact diagonalization data
by less than 4\% (7\%) for $U=8t$ ($20t$) and doping larger than 15\%.
The discrepancy increases when the doping is lowered \cite{Fre92}. It
should also be emphasized that quantitative agreement to quantum Monte
Carlo charge structure factors was established \cite{Zim97}.

The purpose of this paper is to derive and evaluate Fermi liquid
Landau parameters $F_0^a$ and $F_0^s$
for the metallic state in the extended Hubbard model.
Despite of the above discussion on the FM instabilities, the influence
of intersite Coulomb and exchange interactions on Landau parameters
was not analyzed in the Hubbard model until now. We investigate the
instabilities towards uniform FM order and charge instabilities using
the spin rotation invariant (SRI) representation \cite{Li89,FW} of
the Kotliar and Ruckenstein slave boson approach.
We show analytically that weak FM
intersite interactions are sufficient to trigger a divergence in the
magnetic susceptibility expressed by the antisymmetric Landau parameter
$F_0^a$ which implies that the FM instability occurs due to such
interactions for any lattice. For a representative example of a
three-dimensional (3D) cubic lattice we present also numerical analysis,
and compare them at half filling with a two-dimensional (2D) square
lattice.

The paper is organized as follows: The extended Hubbard model is
introduced in Sec. \ref{sec:hub}, together with its Kotliar and
Ruckenstein SRI slave boson representation. In Sec. \ref{sec:saddle} we
perform the saddle-point approximation and present the resulting system
of coupled nonlinear equations. Fluctuations are captured within the
one-loop approximation described in Sec. \ref{sec:loop}, which allows
one to determine analytically Fermi liquid Landau parameter $F_0^a$ at
half filling. Numerical results are presented and discussed in Sec.
\ref{sec:resu}, where we address first $F_0^a$ and the instabilities
towards uniform spin polarization (ferromagnetism) and next consider
$F_0^s$ and charge instabilities in Secs. \ref{sec:f0a} and
\ref{sec:f0s}. The paper is summarized in Sec.~\ref{sec:summa}.

\section{Extended Hubbard model}
\label{sec:hub}

Numerous studies of correlated electrons have been devoted to the
properties of the Hubbard model on a square lattice, especially after
Anderson's proposal that it represents a minimal model for the $d$
electrons within the CuO$_2$ layers common to the high T$_c$
superconductors \cite{And87}. Yet the Hubbard model assumes a perfect
screening of the long-range part of the Coulomb interaction. This may
be questionable and the relevance of this approximation may be
assessed by considering the extended Hubbard model that reads:
\begin{eqnarray}
H &=& \sum_{i,j,\sigma}t_{ij}c_{i\sigma}^{\dagger}c_{j\sigma}
+ U\sum_{i}\left(n_{i\uparrow}-\frac12 \right) \left(n_{i\downarrow}
-\frac12 \right)\nonumber\\
&+&
\frac12 \sum_{i,j}V_{ij}(1-n_{i}) (1-n_{j})
+\frac12 \sum_{i,j}J_{ij} {\vec S_{i}}\cdot{\vec S}_{j},
\label{eq:model}
\end{eqnarray}
and includes intersite Coulomb $V_{ij}$ and exchange $J_{ij}$
interactions. These elements decay fast with increasing distance
$|\vec{R}_i-\vec{R}_j|$, but extend in general beyond nearest
neighbors. Here $c_{i\sigma}^\dagger$ are electron creation
operators at site $i$ with spin $\sigma$,
$n_{i\sigma}=c_{i\sigma}^{\dagger}c_{i\sigma}^{}$, and ${\vec S_i}$ are
spin operators. We consistently use the particle-hole symmetric form
for both density-density interaction terms.

Although one expects that $V_{ij}>0$, in certain cases effective
intersite Coulomb interactions may be attractive \cite{Mic90}.
Therefore, we shall treat $\{V_{ij}\}$ as effective parameters and
consider both signs of them below. Similar, for the exchange elements
$\{J_{ij}\}$ we shall also consider both antiferromagnetic ($J_{ij}>0$)
and ferromagnetic ($J_{ij}<0$) exchange elements. In a transition-metal
oxide the dominating interaction is the superexchange and the
antiferromagnetic coupling is expected. On the contrary, in a metallic
system the direct exchange elements which arise from the Coulomb
interactions could be ferromagnetic. Thus, the present theory makes
predictions concerning itinerant magnetism.

In the original Kotliar and Ruckenstein representation the spin
interaction term remains a four-fermion term. This is no longer the case
in the SRI representation \cite{Li89,FW} which we therefore adopt and
extend for our study. In this framework the Hamiltonian Eq.
(\ref{eq:model}) may be represented as:
\begin{eqnarray}
&H\! &= \sum_{i,j,\sigma}t_{i,j} \sum_{\sigma\sigma'\sigma_{1}}
z^{\dagger}_{i\sigma_{1}\sigma}
f^{\dagger}_{i\sigma}f^{\phantom{\dagger}}_{j\sigma'}
z^{\phantom{\dagger}}_{j\sigma'\sigma_{1}}\nonumber\\
&+& U \sum_{i}\left(d_{i}^{\dagger}d^{\phantom{\dagger}}_{i} -\frac12
   \sum_{\sigma} f^{\dagger}_{i\sigma}f^{\phantom{\dagger}}_{j\sigma'}
   +\frac{1}{4} \right)\nonumber\\
&+&\! \frac{1}{4} \sum_{i,j} V^{\phantom{\dagger}}_{ij}\!\left[\!
\left( 1-\sum_{\sigma} f^{\dagger}_{i\sigma}f^{}_{i\sigma}\!
\right)\!Y_j+Y_i\!\left( 1- \sum_{\sigma}f^{\dagger}_{j\sigma}
f^{\phantom{\dagger}}_{j\sigma} \!\right)\! \right] \nonumber\\
&+&\! \frac12 \sum_{i,j} J^{\phantom{\dagger}}_{ij}
\sum_{\sigma\sigma'\sigma_1}\vec{\tau}_{\sigma\sigma'}
p^{\dagger}_{i\sigma\sigma_1}p^{\phantom{\dagger}}_{i\sigma_{1}\sigma'}
  \cdot \sum_{\rho\rho'\rho_1}\vec{\tau}_{\rho\rho'}
p^{\dagger}_{j\rho\rho_{1}}p^{\phantom{\dagger}}_{j\rho_{1}\rho'}\,,
\nonumber\\
\label{eq:sbmodel}
\end{eqnarray}
where we introduced:
\begin{equation}\label{eqrf:z_SRI}
\underline{z} = e^\dagger
   \underline{L}^{\phantom{\dagger}} M^{\phantom{\dagger}}
   \underline{R}^{\phantom{\dagger}} \underline{p}^{\phantom{\dagger}} +
   \underline{\tilde{p}}^{\dagger} \underline{L}^{\phantom{\dagger}}
   M^{\phantom{\dagger}} \underline{R}^{\phantom{\dagger}}  d \,,
\end{equation}
with $\underline{p}=\frac12\sum_{\mu=0}^3 p_{\mu}\underline{\tau}^{\mu}$
and $\underline{\tau}^{\mu}$ being the Pauli matrices. We further use
$\tilde{p}^{\phantom{\dagger}}_{\sigma\sigma'}\equiv
\sigma\sigma'p^{\phantom{\dagger}}_{-\sigma', -\sigma}$,
and
\begin{eqnarray}
M &=& \left[1+ e^\dagger e + \sum_{\mu} p^\dagger_{\mu}
   p^{\phantom{\dagger}}_{\mu} + d^\dagger d \right]^{\frac12}, \nonumber\\
\underline{L}^{\phantom{\dagger}} &=& \left[
\left(1- d^\dagger d\right) \underline{1}
-2 \underline{p}^{\dagger}\underline{p}^{\phantom{\dagger}}
\right]^{-\frac12}, \nonumber\\
\underline{R} &=& \left[\left(1- e^\dagger e\right) \underline{1}
-2 \underline{\tilde{p}}^{\dagger}\underline{\tilde{p}}^{\phantom{\dagger}}
\right]^{-\frac12} .
\end{eqnarray}
For more details see Ref. \cite{FKW}. In Eq.~(\ref{eq:sbmodel}) we used
the mapping of the fermionic degrees of freedom onto bosons and
expressed the hole doping operator as follows,
\begin{equation}
Y_i\equiv e^{\dagger}_i e^{\phantom{\dagger}}_i
         - d^{\dagger}_i d^{\phantom{\dagger}}_i,
\end{equation}
and the spin operator as
\begin{equation}
{\vec S_i} =  \sum_{\sigma\sigma'\sigma_1}
\vec{\tau}_{\sigma\sigma'}
p^{\dagger}_{i\sigma\sigma_1}p^{\phantom{\dagger}}_{i\sigma_1\sigma'}.
\end{equation}
Since the auxiliary operators span an augmented Fock space, physically
meaningful results may be obtained provided they satisfy local
constraints. For each site they read:
\begin{eqnarray}
\label{eqrf:Qcompl}
e^{\dagger}e+\sum_{\mu} p^{\dagger}_{\mu}p^{\phantom{\dagger}}_{\mu}
+ d^{\dagger}d &=& 1\,, \\
\label{eqrf:Qscal}
\sum_{\sigma} f^{\dagger}_{\sigma} f^{\phantom{\dagger}}_{\sigma} &=&
\sum_{\mu} p^{\dagger}_{\mu}
p^{\phantom{\dagger}}_{\mu}+ 2 d^{\dagger}d\,, \\
\label{eqrf:Qvect}
\sum_{\sigma,\sigma'} f^{\dagger}_{\sigma'} {\vec \tau}_{\sigma\sigma'}
f^{\phantom{\dagger}}_{\sigma} &=& p^{\dagger}_{0} {\vec p}
+ {\vec p}^{\dagger} p^{\phantom{\dagger}}_{0}
-i {\vec p}^{\dagger} \times {\vec p}^{\phantom{\dagger}}\,.
\end{eqnarray}
They may be enforced in path integral formalism.

We remark that
any slave boson representation possesses an internal gauge symmetry
group. In our case it may be made use of to simplify the problem and to
gauge away the phases of the $e$ and $p_{\mu}$ bosons by promoting all
constraint parameters to fields \cite{FW}, leaving us with radial slave
bosons fields \cite{Fre01}. Their approximate values that are obtained
in the saddle-point approximation may be viewed as an approximation to
their exact expectation values that are generically non-vanishing
\cite{Kop07}. The slave boson field corresponding to double occupancy
$d$ has to remain complex however as emphasized by several authors
\cite{Jol91,Kot92,FW}.

Besides, the saddle-point approximation is exact in the large degeneracy
limit, and the Gaussian fluctuations provide the $1/N$ corrections
\cite{FW}. Moreover it obeys a variational principle in the limit of
large spatial dimensions where the Gutzwiller approximation becomes
exact for the Gutzwiller wave function \cite{Met89}. Furthermore, it
could be shown in this limit that longer ranged interactions are not
dynamical and reduce to their Hartree approximation \cite{Mul89}.
Therefore, our approach also obeys a variational principle in this
limit when applied to the above extended Hubbard model Eq.
(\ref{eq:model}).

One can infer from these formal properties
that the approach captures characteristic features of strongly
correlated electrons as the suppression of the quasiparticle residue
and the Mott-Hubbard/Brinkman-Rice transition~\cite{Bri70} to an
insulating state at half filling with increasing on-site Coulomb
interaction. Of particular relevance is the influence of the
longer-ranged Coulomb interaction on the latter transition.

\section{Saddle-point approximation}
\label{sec:saddle}

Though such functional integrals can be calculated exactly for the Ising
chain \cite{Fre01} and some toy models \cite{Kop07}, even with the
Kotliar and Ruckenstein representation \cite{Kop12}, this is unpractical
on higher dimensional lattice. Here we rather resort to the saddle-point
approximation. In the translational invariant paramagnetic phase all
the local quantities are site independent, and the action at
saddle-point reads ($\beta=1/k_BT$),
\begin{equation}\label{eq:spact}
S = \beta L \left(S_B + S_F + \frac{1}{4} U\right)\,,
\end{equation}
with
\begin{eqnarray}
\label{sb}
S_{B}&=&\alpha(e^{2}+d^{2}+p_{0}^{2}-1)-\beta_{0}(p_{0}^{2}+2d^{2})
\nonumber\\
&+&Ud^{2}+ \frac12 V_{0}Y\,, \\
\label{sf}
S_{F}&=&-\frac{1}{\beta}\displaystyle
\sum_{\vec k,\sigma}\ln\left(1+e^{-\beta E_{{\vec k}\sigma}}\right)\,.
\end{eqnarray}
For the extended Hubbard model (\ref{eq:model}) the quasiparticle
dispersion in Eq. (\ref{sf}) reads:
\begin{equation}
E_{{\vec k}\sigma}=z_{0}^{2}t_{\vec k}+\beta_{0}-\frac12 U
-\frac12 V_{0} Y -\mu\;\; ,
\end{equation}
in which we introduced the Fourier transform of the intersite Coulomb
repulsion,
\begin{equation}
\label{eq:V0}
V_{\vec{k}} = \frac{1}{L} \sum_{i,j} V_{ij} e^{-i \vec{k}\cdot
(\vec{R}_j-\vec{R}_i)} \, .
\end{equation}
It is convenient to define the Fourier transform of the intersite
exchange elements in a similar way,
\begin{equation}
\label{eq:J0}
J_{\vec{k}} = \frac{1}{L} \sum_{i,j} J_{ij} e^{-i \vec{k}\cdot
(\vec{R}_j-\vec{R}_i)} \,.
\end{equation}

From Eq. (\ref{eq:spact}) one obtains the following set of saddle-point
equations:
\begin{eqnarray}
\label{eq:speqss}
p_{0}^{2}+e^{2}+d^{2}-1&=&0, \nonumber\\
p_{0}^{2}+2d^{2}&=&n, \nonumber\\
\frac{1}{2e}\frac{\partial z_{0}^{2}}{\partial
   e}\bar{\varepsilon}+\frac12 V_{0}(1-n)\frac{1}{2e}
   \frac{\partial Y}{\partial e}&=&-\alpha, \\
\frac{1}{2p_{0}}\frac{\partial z_{0}^{2}}{\partial p_{0}}
\bar{\varepsilon}&=& \beta_{0}-\alpha, \nonumber\\
\frac{1}{2d}\frac{\partial z_{0}^{2}}{\partial d}\bar{\varepsilon}
+\frac12 V_{0}(1-n)\frac{1}{2d}\frac{\partial Y}{\partial d}&=&
2(\beta_{0}-\alpha)+\alpha-U. \nonumber
\end{eqnarray}
Here we have introduced the averaged kinetic energy,
\begin{equation}
\label{eq:varepsilon}
\bar{\varepsilon}=\int d\omega \rho (\omega)\omega
f_{F}\left(z_{0}^{2}\omega +\beta_{0}-\frac12 U
-\frac12 V_{0} Y -\mu\right)\,,
\end{equation}
which determination involves the density of states $\rho(\omega)$ and
$f_F$ is the Fermi function.
It turns out that the last three equations in the set Eqs.
(\ref{eq:speqss}) may be merged into a single one,
\begin{equation}
\frac{y(1-y)}{n(1-\frac{n}{2})ed}\,\bar{\varepsilon} +U = 0\,.
\end{equation}
Here we have introduced
\begin{equation}
y \equiv (e+d)^2\,,
\end{equation}
in terms of which the saddle-point equation finally reads:
\begin{equation}\label{eq:spfinal}
y^{3}+(u-1)y^{2}=u\delta^{2}\,,
\end{equation}
where $\delta=1-n$ is the doping away from half filling and $u=U/U_{0}$
is the Coulomb parameter in units of $U_{0}$ defined
through
\begin{equation}
U_{0}=-\frac{8}{1-\delta^{2}}\,\bar{\varepsilon}\,.
\end{equation}
At half filling a metal-to-insulator transition occurs for a given
lattice at $U_{c}$, which
is defined as follows
\begin{equation}
U_c\equiv\lim_{\delta \rightarrow 0}U_{0}=-8\bar{\varepsilon}.
\label{uc}
\end{equation}

\begin{figure}[t!]
\psfrag{w}{$\bar{\varepsilon}$}
\begin{center}
\includegraphics[width=8cm]{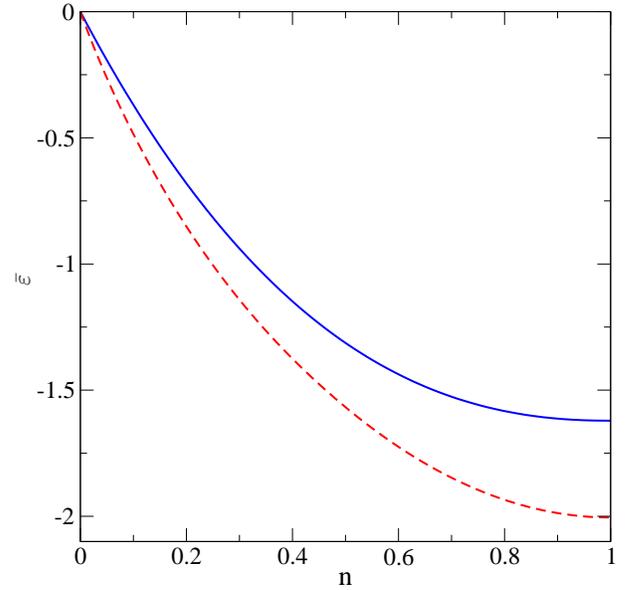}
\end{center}
\caption{(Color online) Average kinetic energy per site
$\bar{\varepsilon}$ for the extended Hubbard model as obtained for the
2D square lattice (blue, solid line) and for the 3D cubic lattice
(red, dashed line) as functions of the electron filling $n$.
Parameters: $t=1$.}
\label{fig:ekin}
\end{figure}

It is very remarkable that neither $J_{ij}$ nor $V_{ij}$ elements enter
Eq. (\ref{eq:spfinal}). Hence, in a paramagnetic phase, the intersite
interactions only influence the fluctuations and do not change electron
localization due to strong onsite interaction $U$. In particular, the
nearest neighbor Coulomb interaction $V$ has no influence on the Mott
gap as discussed by Lavagna \cite{Lav90}. Furthermore the double
occupancy is in exact agreement with the Gutzwiller approximation as
derived by Vollhardt, W\"olfle and Anderson \cite{Vol87}. In the present
case of a 3D cubic lattice the double occupancy vanishes at half filling
for $U_c=16.0387t$ (at the Brinkman-Rice transition \cite{Bri70}).

This behavior is generic and the transition occurs for other lattices in
a qualitatively equivalent way. As an example we take also the case of a
square lattice, where the number of nearest neighbors is reduced from
$z_1=6$ to $z_1=4$. One expects that the value of $U_c$ (\ref{uc}) will
roughly scale with the number of nearest neighbors $z_1$, and corrections
follow from the shape of the density of states. In fact, the averaged
kinetic energy $|\bar{\varepsilon}|$ is reduced somewhat less than by a
factor 2/3 from that found in the cubic lattice due to the shape of the
2D density of states which is closer to a rectangle than in the 3D case,
see Fig. \ref{fig:ekin}. In fact, for the 2D square lattice the
metal-to-insulator transition occurs at $U_c=2 (8/ \pi)^2 t$.

\begin{figure}[t!]
\begin{center}
\includegraphics[width=8.4cm]{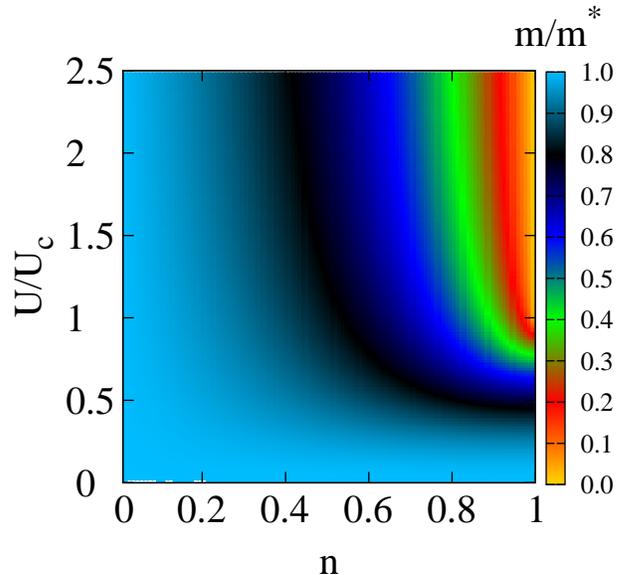}
\end{center}
\caption{(Color online) Inverse effective mass $z_0^2$ (\ref{z2})
for the extended Hubbard model on the cubic lattice. }
\label{fig:z0_I}
\end{figure}

At the critical value of $U=U_c$ the effective mass $m^*$ diverges in
the paramagnetic state (we recall that in the strongly correlated
regime local moments are formed) and its inverse,
\begin{equation}
\label{z2}
z^2\equiv \frac{m}{m^*}\,,
\end{equation}
being quasiparticle residue, vanishes, see Fig. \ref{fig:z0_I}. Here
$m$ is the electron mass and $z^2$ stands for the reduced jump in the
electronic filling, $\langle n_{{\vec k},\sigma}\rangle$, when the
momentum ${\vec k}$ changes from inside to outside of the Fermi surface
\cite{Vol84}. The plot shows that only for filling $n>0.8$ the Fermi
liquid is substantially renormalized in the regime of large $U>U_c$.

Solving the saddle-point equations (\ref{eq:speqss})
yields at half filling,
\begin{equation}
z^2 = 1 - \left(\frac{U}{U_c}\right)^2\,.
\end{equation}
The present analysis shows that its doping dependence is universal in
the extended Hubbard model, and is influenced neither by $V_{ij}$ nor
by $J_{ij}$.

\section{One-loop approximation to the response functions}
\label{sec:loop}

Having mapped all degrees of freedom onto bosons allows us for directly
evaluating the spin and charge response functions. Indeed, following
Ref. \cite{Zim97}, the spin and
density fluctuations may be expressed as
\begin{eqnarray}
\delta S_z &\equiv & \sum_{\sigma} \sigma \delta n_{\sigma} =
\delta (p_0^{\dagger} p^{\phantom{\dagger}}_z
+ p_z^{\dagger}p^{\phantom{\dagger}}_0), \\
\delta N &\equiv & \sum_{\sigma}\delta n_{\sigma} =
\delta (d^{\dagger} d-e^{\dagger} e),
\end{eqnarray}
in the SRI representation. The spin and charge autocorrelation functions
can be written in terms of the slave boson correlation functions as:
\begin{eqnarray}
\label{korrs}
\chi_{s}(k) \!& = &\sum_{\sigma,\sigma^{'}}\sigma\sigma^{'}
\langle \delta n_{\sigma}(-k)\delta n_{\sigma^{'}}(k)\rangle
  = \langle \delta S(-k) \delta S(k)\rangle, \nonumber \\
\label{korrn}
\chi_{c}(k)\! & = &\sum_{\sigma\sigma^{'}}
\langle \delta n_{\sigma}(-k)\delta n_{\sigma^{'}}(k)\rangle
  = \langle \delta N (-k) \delta N(k)\rangle. 
\end{eqnarray}
Here we use the notation $k\equiv(\vec{k},\omega)$.
Performing the calculation to one-loop order, one can make use of the
propagator $S_{ij}(k)$ as given in the Appendix. The susceptibilities
are given by the matrix elements of the inverse propagator as follows:
\begin{eqnarray}
\label{korrend}
\chi_{s}(k) & = & 2p_{0}^{2}S_{77}^{-1}(k),
\nonumber \\
\label{korrcc}
\chi_{c}(k) & = &
2e^{2}S_{11}^{-1}(k)-4edS_{12}^{-1}(k)+2d^{2}S_{22}^{-1}(k).
\end{eqnarray}
As emphasized and analyzed by several authors, see e.g. Refs.
\cite{Li91,LiBen}, the Fermi liquid behavior is obtained when
considering the above $\chi_{s}(k)$ and $\chi_{c}(k)$ in the long
wavelength and low frequency limit. However, in contrast to the
conventional random phase approximation (RPA) results, the obtained
Landau parameters involve effective interactions,
which differ in the spin channel and in the charge channel.

Explicitly the spin susceptibility reads:
\begin{equation} \label{chis}
\chi_{s}(k)=\frac{\chi_{0}(k)}
{1 +\!A_{\vec k}\chi_{0}(k) +\!B\chi_{1}(k) +\!
C[\chi_{1}^{2}(k) -\!\chi_{0}(k)\chi_{2}(k)]},
\end{equation}
where
\begin{eqnarray}\label{theas}
A_{\vec k} & = & \frac{
\alpha-\beta_{0}+\varepsilon_{0}z_{0}
\frac{\partial^{2}z_{\uparrow}}{\partial p_{3}^{2}}+
\varepsilon_{\vec k}\left(\frac{\partial z_{\uparrow}}
{\partial p_{3}}\right)^{2}}{2p_{0}^{2}} + \frac14 J_{\vec k}, \\
B & = & \frac{z_{0}}{p_{0}}\left(\frac{\partial
z_{\uparrow}}{\partial p_{3}}\right)\,, \\
C & = & \left(\frac{z_{0}}{2 p_{0}}\right)^{2}
\left(\frac{\partial z_{\uparrow}}{\partial p_{3}}\right)^{2}.
\end{eqnarray}

Using
\begin{eqnarray}\label{dzup}
\frac{\partial z_{\uparrow}}{\partial p_{3}}& = & \sqrt{2}\,
\frac{(e-d)(1-\delta^2) -2 p_0^2 \delta(e+d)}{(1-\delta^2)^{3/2}}\,, \\
\frac{\partial^{2}z_{\uparrow}}{\partial p_{3}^{2}} & = &
\frac{\sqrt{2}p_0}{(1-\delta^2)^{5/2}} \left\{2(e+d)
\left[1-\delta^2 + 2p_0^2(1+2\delta^2)\right]\right.  \nonumber \\
&& - \left. 4 \delta (e-d) (1-\delta^2) \right\}\,.
\end{eqnarray}
together with the saddle-point Eqs. (\ref{eq:speqss}) yields the
Landau parameter $F^a_0$ at half filling as:
\begin{equation}\label{eq:F0aHub}
F^a_0= 2 \bar{\varepsilon} N_F^{(0)} \frac{u(1-u)(2+u) + (1+u)
(J_{0}/8\bar{\varepsilon})}{(1+u)^2(1-u)}\,,
\end{equation}
where we have introduced the bare density of states at the Fermi energy
$N_F^{(0)}\equiv\rho(E_F)$. We decompose Eq. (\ref{eq:F0aHub}) into a
regular and singular part, and using Eq. (\ref{uc}) we find,
\begin{equation}
\label{eq:F0a}
F^a_0= 2 N_F^{(0)}\bar{\varepsilon}\left\{ \frac{u(2+u)}{(1+u)^2}\,
-\frac{J_0/U_c}{1-u^2}\,\right\}\,.
\end{equation}
The latter term depends solely on the $\vec{k}=0$ component of the
exchange coupling, $J_0\equiv J_{\vec{k}=0}$ defined in Eq.
(\ref{eq:J0}). We emphasize that the ferromagnetic instability
deduced from Eq. (\ref{eq:F0a}) is general and occurs in all cases
\textit{below} the metal-insulator transition when $J_0<0$. This
remarkable result follows from the band narrowing when $U\to U_c$
which amplifies the effects of the intersite exchange interaction.

Moreover, spin and charge fluctuations separate at the one-loop order,
and the intersite Coulomb elements in $V_{ij}$ (\ref{eq:V0}) have
no effect on the value of $F^a_0$. Unfortunately, a similar analytic
result for $F^s_0$ could not be derived at half filling as the
dependence on $V_0\equiv V_{\vec{k}=0}$ defined in Eq. (\ref{eq:V0})
is not sufficiently transparent.

Equation (\ref{eq:F0a}) is the central analytic result obtained at half
filling. Its physical origin lies in the fact that, in the limit of
vanishing hopping, the Hubbard model favors the formation of localized
magnetic moments that order according to the exchange couplings, for
instance ferromagnetically for $J_0 <0$. Our result indicates that a
minimum of coherence of the quasi-particles $z_F^2$ is necessary to
destabilize the ferromagnetic order. It only depends on
$j_0 \equiv J_0/U$ and, for a rectangular density of states (DOS), reads:
\begin{equation}
\label{eq:ZF}
z_F^2 = \frac{4 j_0 + j_0^2 + (1-j_0) \sqrt{1-6 j_0+ j_0^2} -1}{4j_0^2}\,,
\end{equation}
which behaves as $z_F^2 \simeq -2 j_0$ for small ferromagnetic exchange.
Hence, for $J_0 \rightarrow 0^{-}$ the ferromagnetic instability takes
place at $U=U_c$, while it is absent for $J_0 =0$. As a result we have
shown that introducing an arbitrarily weak exchange $J_0\ne 0$
interaction results in a singular behavior of $F^a_0$ in the vicinity
of the Brinkman-Rice point $U_c$ ($u=1$). This is in marked contrast to
the known results for a flat band in the absence of intersite exchange:
\begin{equation}
\label{eqrf:f0a}
F^a_0  = -1 + \frac{1}{(1+U/U_c)^2}\,.
\end{equation}
While the simple form
\begin{equation}
\label{eqrf:f0s}
F^s_0  =  \frac{U(2U_c-U)}{(U_c-U)^2}=
-1 + \frac{1}{(1-U/U_c)^2}\,,
\end{equation}
could be derived in absence of Coulomb elements this is no longer
the case when they are taken into account. Note that the property
$F^s_0(U)=F^a_0(-U)$ can be derived on a more general ground
\cite{Vol84}.

\section{Numerical results}
\label{sec:resu}

\subsection{Ferromagnetic instability --- $F_0^a$ parameter}
\label{sec:f0a}

\begin{figure}[b!]
\begin{center}
\includegraphics*[width=8.2cm ]{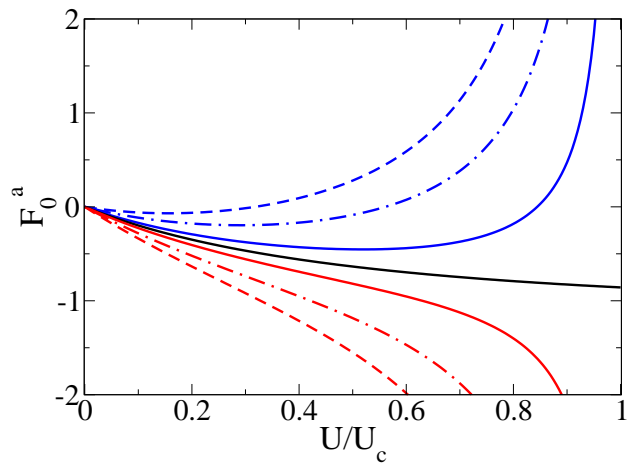}
\end{center}
\caption{(Color online)
Landau parameter $F^a_0$ for the extended Hubbard model at half filling
on the cubic lattice. Different lines from top to bottom for decreasing
$J/U=0.2$, 0.1, 0.04, 0, $-0.04$, $-0.1$, $-0.2$.
}
\label{fig:F0ahf}
\end{figure}

In this Section we compute the Landau parameter $F^a_0$ for the
extended Hubbard model with nearest neighbor exchange or Coulomb
interactions on the cubic lattice to analyze its filling dependence.
We first investigate half
filling ($n=1$), evaluating the obtained analytic expression Eq.
(\ref{eq:F0a}) to demonstrate the divergent behavior at finite intersite
exchange. The Hubbard model does not exhibit the FM instability in the
metallic regime as $F^a_0>-1$, see Fig.~\ref{fig:F0ahf}. However, as
anticipated above, an infinitesimal FM coupling $J_0$ generates an
instability of the nonmagnetic state at half filling when the Coulomb
interaction $U$ approaches $U_c$.

\begin{figure}[t!]
\begin{center}
\includegraphics*[width=9.0cm]{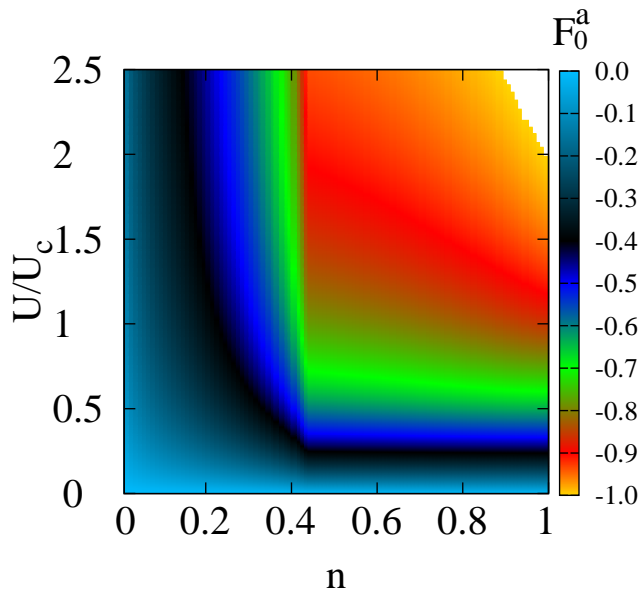}
\end{center}
\caption{(Color online)
Landau parameter $F^a_0$ for the Hubbard model on the cubic lattice.
The FM instability occurs only near half filling (at the boundary of
the white region where $F^a_0<-1$) at very large $U>32t$ ($U>2U_c$).
}
\label{fig:F0a_0}
\end{figure}

\begin{figure}[b!]
\begin{center}
\includegraphics*[width=8.2cm ]{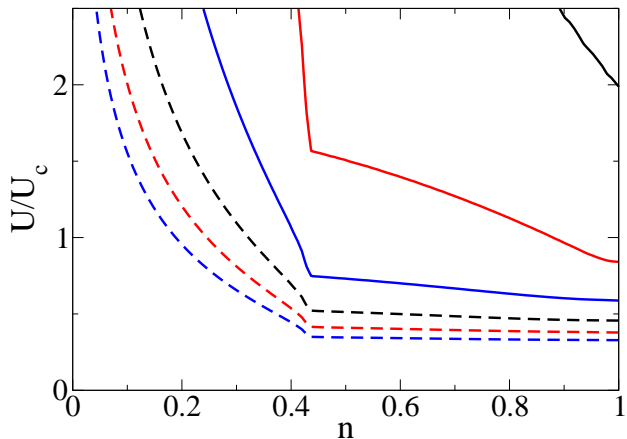}
\end{center}
\caption{(Color online)
Instability lines of the unpolarized state towards FM order as given
by the divergence of the magnetic susceptibility (Landau parameter
$F^a_0=-1$) for the extended Hubbard model with FM exchange $J<0$ on
the cubic lattice. Different lines from top to bottom for:
$J/U=0$, $-0.01$, $-0.05$, $-0.1$, $-0.15$, $-0.2$. }
\label{fig:F0a_mp1}
\end{figure}

Although Eq. (\ref{eq:F0a}) is more general, we shall consider below
the extended Hubbard model with nearest neighbor exchange interactions
only,
\begin{equation}
\label{eq:Jij}
J_{ij}=\left\{
\begin{array}{ r l}
J & \mbox{for a bond $\langle ij\rangle$ with $j\in{\cal N}(i)$}  \\
0 & \mbox{otherwise}
\end{array}
\right.,
\label{eq:J}
\end{equation}
where ${\cal N}(i)$ stands for the set of nearest neighbors of site $i$.
We have found that the model is unstable at half filling against the FM
order near the metal-to-insulator transition. The location of the
instability depends rather sensitively on the FM coupling, from $U_c$
for $J=0^+$ down to $0.33U_c$ for $J/U=-0.2$. Larger values of $-J/U$
are nonphysical and they are not considered below. On the contrary, a
positive (AF) intersite exchange coupling $J_0>0$ suppresses the
tendency towards ferromagnetism, and the parameter $F^a_0$ becomes
positive and large near the metal-to-insulator transition, see Fig.
\ref{fig:F0ahf}.

\begin{figure}[t!]
\begin{center}
\includegraphics*[width=9.0cm ]{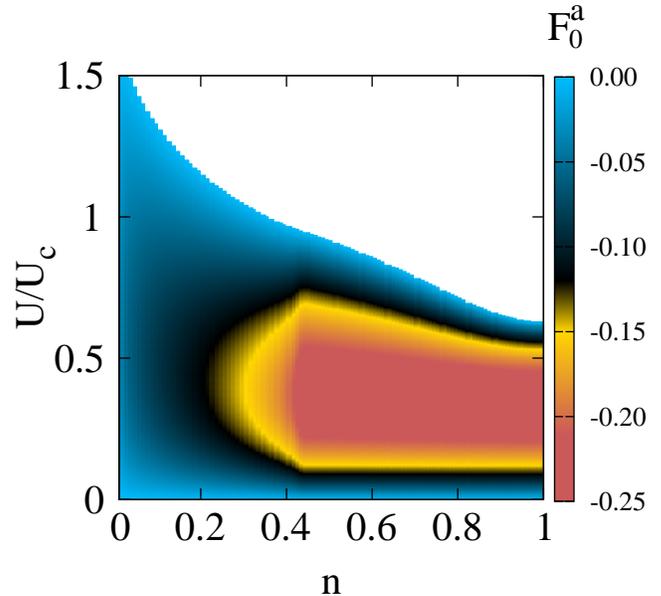}
\end{center}
\caption{(Color online)
Landau parameter $F^a_0$ for the extended Hubbard model with AF $J/U=0.1$
on the cubic lattice. At sufficiently large $U$ the parameter $F^a_0$ is
positive (white region). }
\label{fig:F0a_p1}
\end{figure}

Since the Landau parameter $F^a_0$ is directly proportional to the
product of the bare kinetic energy and the density of states at $E_F$,
$\bar{\varepsilon}N_F^{(0)}$, it shows a lesser sensitivity to the
band structure. For instance, one finds at half filling
$\bar{\varepsilon}N_F^{(0)}=-0.5$
for a flat (rectangular) DOS, and
$\bar{\varepsilon}N_F^{(0)}\simeq -0.57$
for the cubic lattice.
This gives the value of $F_0^a\simeq -0.86$ at $J=0$. We have verified
that similar values are obtained for other typical DOSs
at half filling, except for the 2D square lattice, where the van Hove
singularity at $n=1$ results in the FM instability at infinitesimal
$U=0^+$ \cite{Fle97} - disregarding the stronger AF instability
following from nesting.

Next we consider the doping dependence of the Landau parameter $F^a_0$
in the Hubbard model, i.~e., at $J_{ij}=0$. As one can see in Fig.
\ref{fig:F0a_0}, $F^a_0$ reaches the critical value $F^a_0=-1$ for
$U\simeq 32t$ at $\delta=0^+$, namely in the regime where Nagaoka
ferromagnetism is expected. The critical $U$ increases linearly with
doping. Furthermore, the cusp in the density of states for the cubic
lattice at $n\simeq 0.43$ makes itself clearly visible in Fig.
\ref{fig:F0a_0}, with reduced values of $|F^a_0|$ at lower electron
density.

Also away from half filling, finite FM exchange coupling $J_0<0$
triggers the FM instability $F^a_0=-1$ at significantly lower values of
$U$. For instance, already $J/U=-0.01$ brings this instability down to
the values of $U\sim 20t$ for the doping $\delta<0.57$
(filling $n>0.43$) where the DOS is almost independent of energy,
see Fig.~\ref{fig:F0a_mp1}. When $J/U=-0.05$, the FM instability occurs
at $U<10t$ in the same doping regime, and comes down also for the lower
electron fillings. For lower $J_0$ the FM instability occurs at even
lower values of $U$. This is in contrast to the calculations to the
two-band model, where the FM instability was only found in the doped
Mott insulator regime \cite{Lamb}. In that case, no intersite FM
coupling is needed and the FM instability follows from Hund's exchange.

On the contrary, an AF coupling suppresses the FM instability, and the
value of $J_0=0.1$ totally removes ferromagnetism
as shown in Fig.~\ref{fig:F0a_p1}. For the small values of $U$ the Landau
parameter $F^a_0$ is negative in the entire regime of $n$, but then
changes sign when $U$ approaches $U_c$. The exception here is the regime
of low filling $n<0.2$, where $F^a_0<0$ for $U<U_c$. Larger AF exchange
coupling $J_0/U=0.2$ leaves only a narrow range of $F^a_0<0$ (not shown).

\subsection{Charge instability --- $F_0^s$ parameter}
\label{sec:f0s}

\begin{figure}[t!]
\begin{center}
\includegraphics*[width=9.0cm ]{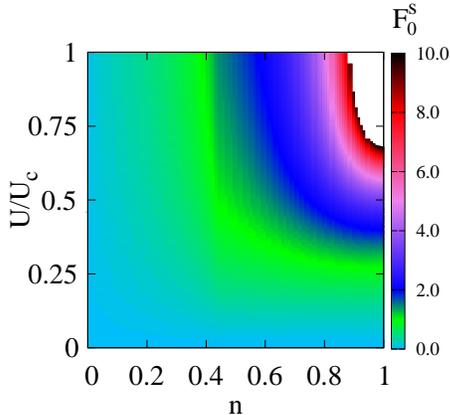}
\end{center}
\caption{(Color online)
Landau parameter $F^s_0$ for the Hubbard model on the cubic lattice.
Here the white region stands for values $F^s_0>10$. No instability
is found.}
\label{fig:f0s}
\end{figure}

The symmetric Landau parameter $F^s_0$ which stands for the charge
response has to be evaluated numerically even at half filling, except
for $V=0$. As expected, $F^s_0$ vanishes both for $U=0$ and for $n=0$,
as $F^a_0$ does. Otherwise, unlike the antisymmetric Landau parameter
$F^a_0$ which decreases with increasing $U$ in the Hubbard model
(at $V_0=0$), the symmetric parameter $F^s_0$ increases with $U$ in the
entire regime of filling $0<n\leq 1$.
This increase is stronger near half filling, where $F^s_0>10$ for
$U/U_c>0.7$ in a range of small doping away from half filling, see Fig.
\ref{fig:f0s}. At half filling the value of the positive $F^s_0$ is
given by Eq. (\ref{eqrf:f0s}). It rapidly increases and finally
diverges at the metal-to-insulator transition (we recall that for the
simple cubic lattice $U_c\simeq 16.04t$).
Away from $n=1$ the increase of $F^s_0$ is moderate and it follows the
same pattern as $1/z^2$ in Fig. \ref{fig:z0_I}, being another
manifestation of strong electron correlations near half filling.

\begin{figure}[b!]
\begin{center}
\includegraphics*[width=9.0cm ]{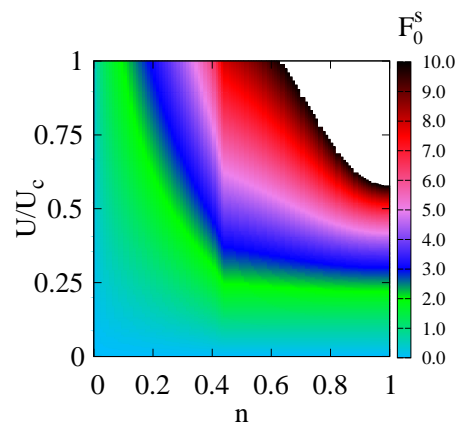}
\end{center}
\caption{(Color online)
Landau parameter $F^s_0$ for the extended Hubbard model on the cubic
lattice. Increase of $F^s_0$ with increasing $U/U_c$ is enhanced by
positive $V/U=0.2$; large values of $F^s_0>10$ are found for $n>0.6$.}
\label{fig:f0s_p2}
\end{figure}

The increase of $F^s_0$ with increasing $U/U_c$ is enhanced by a
positive intersite Coulomb repulsion in the extended Hubbard model.
Here we also consider again the case of nearest neighbor
interactions,
\begin{equation}
\label{eq:Vij}
V_{ij}=\left\{
\begin{array}{ r l}
V & \mbox{for a bond $\langle ij\rangle$ with $j\in{\cal N}(i)$}  \\
0 & \mbox{otherwise}.
\end{array}
\right.,
\label{eq:V}
\end{equation}
where we use the same notation as in Eq. (\ref{eq:Jij}). This case is
particularly favorable for a charge instability near half filling, for
instance in the form of a checkerboard state stable for large values of
$V$ as then the cost of any intersite Coulomb energy $\propto V$ can be
avoided. However, such a state does not occur as a consequence of the
investigated instability of the uniform state. When $V>0$,
one finds even a stronger increase of $F^s_0$ near half filling, and
finally it becomes even larger than $F^s_0=10$ in a broad range of
filling $n>0.6$ for the cubic lattice at $V=0.2U$, see Fig.
\ref{fig:f0s_p2}. The uniform charge distribution is therefore more
robust in the regime of $n\simeq 1$, if $V/U>0$.

\begin{figure}[t!]
\begin{center}
\includegraphics*[width=9.0cm ]{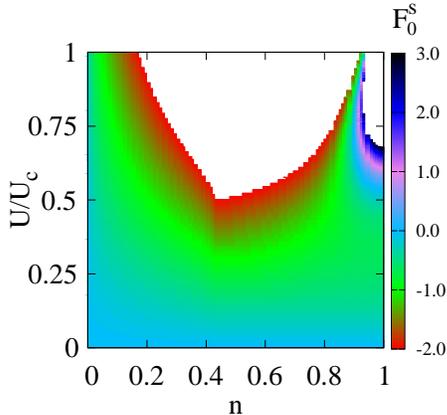}
\end{center}
\caption{(Color online)
Landau parameter $F^s_0$ for the extended Hubbard model on the cubic
lattice with attractive intersite interaction $V/U=-0.2$. Large values
of $F^s_0>3$ are found only near $n=1$, while the charge instability
$F^s_0=-1$ occurs for a broad range of $0.045<n<0.93$. Note that the
instability line $F^s_0=-1$ extends to $n=1^-$, and stops
at $U\simeq 1.246 U_c$.}
\label{fig:f0s_m2}
\end{figure}

We have found that the uniform charge distribution is destabilized by
attractive charge interactions $V<0$, particularly in the regime near
quarter filling $n\simeq 0.5$. At $V=-0.2U$ the value of $F^s_0$
decreases with increasing $U$ for any filling $n$ and this decrease is
fastest near quarter filling. For $U<U_c$ one finds the charge
instability at $F^s_0=-1$ in a broad range of $n\in(0.045,0.93)$.
This instability is related to the shape of the DOS and is easiest to
realize at $n\simeq 0.42$, there the DOS has a van Hove singularity.

\begin{figure}[t!]
\begin{center}
\includegraphics[width=8.0cm]{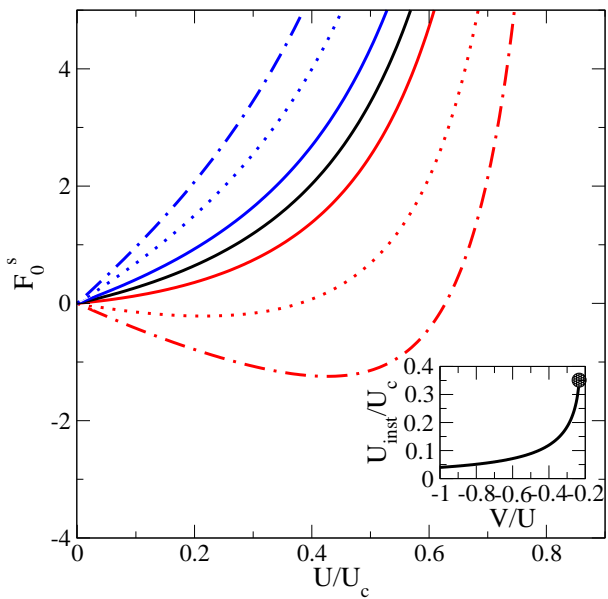}
\end{center}
\caption{(Color online)
Landau parameter $F^s_0$ for the extended Hubbard model on the cubic
lattice at half filling ($n=1$) for selected decreasing values of
intersite Coulomb interaction $V$ from top to bottom:
V/U=0 (black line),
$V/U>0$ (blue) --- $V/U=0.05$ (solid line),
$V/U = 0.15$ (dotted) and $V/U = 0.25$ (dashed-dotted line), and
$V/U<0$ (red) --- $V/U=-0.05$ (solid line),
$V/U =-0.15$ (dotted) and $V/U =-0.25$ (dashed-dotted line).
The inset shows the instability value $U_{\rm inst}/U_c$ for
$V/U\in[-1.0,-0.2]$. Its end point is marked by a solid circle.
}
\label{fig:n=1}
\end{figure}

The data of Figs. \ref{fig:f0s_p2} and \ref{fig:f0s_m2} suggest that
in the case of charge response the regime near the metal-insulator
transition at half filling is robust and the Landau parameter $F^s_0$
is here always enhanced, even for $V<0$. This finding confirms that
electrons are strongly correlated and the system properties change
radically at the metal-to-insulator transition at $U_c$. Hence we
inspect now the case $n=1$ in more detail. We remark that
the effect of finite $V$ first resembles somewhat that of finite $J$
for weak to moderate coupling: $F^s_0$ is reduced for attractive $V$
while it is enhanced for repulsive $V$, see Fig. \ref{fig:n=1}. The
reduction of $F^s_0$ occurs only for sufficiently large $-V$ and is
visible in Fig. \ref{fig:n=1} for $V/U=-0.15$, and beyond. As a result
a minimum in $F^s_0$ develops at $U\simeq 0.4U_c$, the minimal value
of $F^s_0$ decreases with increasing $-V$, and a charge instability may
be found at the critical value $V/U<-0.234$, see the inset in Fig.
\ref{fig:n=1}. The instability moves towards lower values of $U$ with
decreasing $V$ when the minimum of $F^s_0$ becomes deeper with further
decreasing $V$. Particularly interesting is the non-monotonic behavior
of $F^s_0$ with increasing $U$ for $V<0$. We therefore suggest that a
sufficiently strong intersite Coulomb attraction $-V/U>0.234$ is
necessary to induce phase separation. The instability is absent for
repulsive $V$, where the uniform charge distribution is locally stable.

\section{Discussion and summary}
\label{sec:summa}

We have presented the consequences of finite intersite interactions:
exchange $J_{ij}$ and Coulomb $V_{ij}$, and their impact on the
instabilities of correlated electrons on a cubic lattice in the
nondegenerate band described by the extended Hubbard model. While the
Hubbard model was originally introduced, \textit{inter alia}, to explain
the metallic ferromagnetism, this idea fails as the model does not
predict an instability of the metallic phase towards spin polarized
(weakly ferromagnetic) states for the majority of lattices. Such an
instability is expected in first place at half filling where electron
correlation effects are strongest, so this case is of particular
significance. However, it is well known that the metal-to-insulator
transition happens typically for a \textit{lower} value of $U$ than
the ferromagnetic instability, given by the divergence of spin
susceptibility when the antisymmetric Landau parameter takes the value
$F^a_0=-1$.

The analytic result derived in this paper shows that intersite
exchange interactions change radically the results for the metallic
phase below the metal-to-insulator transition near $U_c$. If these
interactions are globally ferromagnetic, a spin-polarized ground state
occurs here before the system undergoes the metal-to-insulator
transition, i.~e., at $U<U_c$. This result could be of importance to
decide about the properties of some compounds with strongly correlated
electrons, for instance, heavy fermion systems.

Somewhat surprisingly we have found that the symmetric (charge)
Landau parameter
$F^s_0$ has a rather rich behavior when intersite Coulomb interactions
are included. Small interactions do not change the value of $F^s_0$
significantly. Most unexpected is the absence of a charge instability
at half filling for moderate intersite Coulomb interactions,
independently of the sign of the intersite Coulomb interaction, as it
takes a sufficiently large attractive Coulomb interactions $-V/U>0.234$
for a charge instability to occur. For smaller $-V$, it develops in the
intermediate regime of doping, i.~e., for $0.045<n<0.93$. It may be
expected that the intersite Coulomb elements are repulsive
\cite{Han11,Han13}, except for systems with strong polaronic effects,
which gives no charge instability. Whether or not such interactions
could be sufficiently attractive to cause an instability is an
experimental challenge for future studies.

Summarizing, we have shown that the Landau parameters $F^a_0$ and
$F^s_0$ are both sensitive to different intersite elements --- $F^a_0$
is modified only by exchange elements, while  $F^s_0$ only by Coulomb
elements.
The present theory predicts a ferromagnetic instability in a strongly
correlated metallic system with globally ferromagnetic exchange.
The derived analytic result for $F^a_0$ Eq. (\ref{eq:F0a}) is very
remarkable as it uncovers that for any lattice the Hubbard model at
half filling is at the verge of the ferromagnetic instability, which
is triggered by an infinitesimal ferromagnetic intersite exchange.
This result provides a new context for the original idea of Kanamori
\cite{Kan63} who introduced the Hubbard model as the simplest model
of itinerant ferromagnetism.

At the same time, interesting behavior of the symmetric Landau
parameter $F^s_0$ was found for attractive nearest neighbor Coulomb
interactions. In contrast to $F^a_0$, a more interesting result is
found here away from half filling, where attractive interactions lead
to charge instabilities signaling a tendency towards phase separation.
We also remark that we investigated here only instabilities towards
ferromagnetism or phase separation. Other instabilities may also
occur at finite values of $\vec{k}$ which is an interesting subject
for future research.

\acknowledgments

We thank Thilo Kopp, Philipp Hansmann, Peter Horsch, and Olivier
Juillet for insightful discussions.
R.F. is grateful for the warm hospitality at MPI Stuttgart where part
of this work has been done, and to the R\'egion Basse-Normandie and
the Minist\`ere de la Recherche for financial support. A.M.O. kindly
thanks Universit\'e de Caen Basse-Normandie for financial support and
Laboratoire CRISMAT for the warm hospitality, where part of this work
was done. He also acknowledges
support by Narodowe Centrum Nauki (NCN, National Science Center)
under Project No. 2012/04/A/ST3/00331.

\appendix*

\section{Inverse propagator matrix}

The derivation of the inverse propagator matrix for lattices with
inversion symmetry follows the one by
Zimmermann et al. \cite{Zim97}. To second order in the fluctuations
the action $S$ may be decomposed into two parts:
\begin{equation}
S = S^{({\rm spin})} + S^{({\rm charge})},
\end{equation}
where
\begin{equation}
S^{({\rm charge})} = \sum_{q,\mu,\nu}
\delta \psi_{\mu}(-q)S_{\mu,\nu}^{C}(q)\delta \psi_{\nu}(q)\,,
\end{equation}
with $1\leq \mu\ (\nu) \leq 6$, and
\begin{equation}
S^{({\rm spin})} = \sum_{q,\mu,\nu}
\delta \psi_{\mu}(-q)S_{\mu,\nu}^{S}(q)\delta \psi_{\nu}(q)\,,
\end{equation}
with $7\leq\mu\ (\nu)\leq 12$ including the three spin components.
We use the following sequence of boson fields:
$\psi_{1}=e$, $\psi_{2}=d'$, $\psi_{3}= d''$, $\psi_{4}=p_0$,
$\psi_{5}=\beta_0$, $\psi_{6}=\alpha$, $\psi_{7}=p_x$, $\psi_8=\beta_x$,
$\psi_9=p_y$, $\psi_{10}=\beta_y$, $\psi_{11}=p_z$, $\psi_{12}=\beta_z$,
after having introduced the short-hand notation $d' = \Re{(d)}$ and
$d''= \Im{(d)}$.

Regarding the spin sector at zero frequency we find:
\begin{eqnarray}
S_{7,7}({\vec k}) & = & S_{9,9}({\vec k})=S_{11,11}({\vec k}) \nonumber \\
&\equiv&\alpha-\beta_{0}+\frac12 p_0^2 J_{\vec k},
\nonumber \\
&+&\varepsilon_{0} z_{0}\frac{\partial^{2} z}
{\partial p_z^2}+
\left[\varepsilon_{\vec k}-\frac{1}{2}z_{0}^{2}\chi_{2}({\vec k})\right]
(\frac{\partial z_{\uparrow}}{\partial p_z})^2 \, ,\nonumber \\
S_{8,8}({\vec k}) & = & S_{10,10}({\vec k})=S_{12 ,12}({\vec k})\equiv\
-\frac12 \chi_0({\vec k}),
\nonumber \\
S_{7,8}({\vec k}) & = & S_{9,10}({\vec k})=S_{11,12}({\vec k})
\nonumber \\
&\equiv& -p_{0} - \frac12 z_0 \chi_{1}({\vec k}) \frac{\partial z_{\uparrow}}
{\partial p_z}.
\end{eqnarray}
Here we used the Fourier transform $J_{\vec k}$ Eq. (\ref{eq:J0}) of
the intersite exchange elements. Notably, regarding the spin sector,
the only difference to the Hubbard model \cite{Zim97} is the presence
of $J_{\vec k}$ in $S_{7,7}({\vec k})$.

Regarding the charge sector  at zero frequency we obtain $S^{C} = S + \tilde
S^{(I)}$ with the non-vanishing matrix elements of the symmetric matrix $S$:
\begin{eqnarray}
S_{1,1}({\vec k}) & = & \alpha+\frac12 V_{0}\,,
\nonumber \\
S_{2,2}({\vec k}) & = & \alpha-2\beta_{0}+U-\frac12 V_{0}\,,
\nonumber \\
S_{3,3}({\vec k}) & = & \alpha-2\beta_{0}+U-\frac12 V_{0}\,,
\nonumber \\
S_{4,4}({\vec k}) & = & \alpha-\beta_{0}\,,
\nonumber \\
S_{1,6}({\vec k}) & = & e\,,
\nonumber \\
S_{2,5}({\vec k}) & = & -2d\,,
\nonumber \\
S_{2,6}({\vec k}) & = & d\,,
\nonumber \\
S_{4,5}({\vec k}) & = & -p_{0}\,,
\nonumber \\
S_{4,6}({\vec k}) & = & p_{0}\,.
\end{eqnarray}
We introduced above
$\tilde S^{(I)}({\vec k})=\tilde S({\vec k})+\tilde S^{(V)}({\vec k})$,
with the non-vanishing matrix elements of $\tilde S({\vec k})$ defined as:
\begin{eqnarray}
\label{sti}
\tilde S_{\mu,\nu}({\vec k}) &=& \varepsilon_{0} z_{0}
\frac{\partial^{2}z}{\partial \psi_{\mu} \partial \psi_{\nu}}+
\left(\varepsilon_{\vec k} -\frac12 z_{0}^{2} \chi_2 ({\vec k})\right)
\frac{\partial z}{\partial \psi_{\mu}}
\frac{\partial z}{\partial \psi_{\nu}}\,,
\nonumber\\
&& \mu, \nu=1,2,4 \nonumber\\
\tilde S_{3,3}({\vec k}) &=& \varepsilon_{0} z_{0}
\frac{\partial^{2}z}{\partial d''^{2}}+
\left(\varepsilon_{\vec k} -\frac12 z_{0}^{2} \chi_2' ({\vec k})\right)
\frac{\partial z}{\partial d''} \frac{\partial z^{*}}{\partial d''}\,,
\nonumber\\
\tilde S_{\mu,5}({\vec k}) & = & -\frac{1}{2} z_{0} \frac{\partial z}
{\partial\psi_{\mu}} \chi_{1}({\vec k}) \,, \nonumber\\
&& \mu=1,2,4 \nonumber\\
\tilde S_{5,5}({\vec k}) & = & -\frac{1}{2} \chi_{0}({\vec k}) .
\end{eqnarray}
These elements are defined in terms of:
\begin{eqnarray}
\label{eqchin}
\chi_{n}({\vec k})&=& -\sum_{\vec{p},\sigma}
(t_{\vec{p}} + t_{\vec{k} + \vec{p}})^n
\frac{f_F(E_{\vec{k} + \vec{p}}) - f_F(E_{\vec{p}})}
{E_{\vec{k} + \vec{p}} - E_{\vec{p}}}\,, \nonumber\\
\chi_{2}'({\vec k})&=& -\sum_{\vec{p},\sigma}
(t_{\vec{p}} - t_{\vec{k} + \vec{p}})^2
\frac{f_F(E_{\vec{k} + \vec{p}}) - f_F(E_{\vec{p}})}
{E_{\vec{k} + \vec{p}} - E_{\vec{p}}}\,, \nonumber\\
\varepsilon_{\vec k}&=& \sum_{{\vec p},\sigma} t_{\vec{p}
- \vec{k}}f_F(E_{\vec{p}}).
\end{eqnarray}
For ${\vec k}=0$, the relation $\chi_{n}(0) = (2t_F)^n \chi_{0}(0)$ may be
established \cite{LiBen}.

Explicitly, using $V_{-\vec k} = V_{\vec k}$ due to the inversion
symmetry of the lattice, the relevant matrix elements of the symmetric
matrix $\tilde{S}^{(V)}$ read:
\begin{eqnarray}
\tilde{S}^{(V)}_{1,1}({\vec k})\! &\!=\!& \frac12\! \left[ -e^2 V_{\vec
k}^2\chi_{0}({\vec k})
   +2e z_0\frac{\partial z}{\partial e} V_{\vec k} \chi_{1}({\vec k})
-\varepsilon_{V}\right]\, ,\nonumber\\
\tilde{S}^{(V)}_{1,2}({\vec k})\! &\!=\!& \frac12\! \left[ e d V_{\vec
k}^2\chi_{0}({\vec k})
+\! z_0 V_{\vec k} \chi_{1}({\vec k})\left(e\frac{\partial z}{\partial d'}
\!-\!d  \frac{\partial z}{\partial e} \right)
\right]\!,\nonumber\\
\tilde{S}^{(V)}_{1,4}({\vec k})\! &\!=\!& \frac12\,e z_0 V_{\vec k}
\frac{\partial z}{\partial p_0} \chi_{1}({\vec k})\,, \nonumber\\
\tilde{S}^{(V)}_{1,5}({\vec k})\! &\!=\!& \frac12\,e\,V_{\vec k}\,
\chi_{0}({\vec k})\,,
\nonumber\\
\tilde{S}^{(V)}_{2,2}({\vec k})\! &\!=\!& \frac12\!
\left[ -d^2 V_{\vec k}^2\chi_{0}({\vec k})
-\!2 d z_0 \frac{\partial z}{\partial d'} V_{\vec k}\chi_{1}({\vec k})
+ \varepsilon_{V}\right]\!,\nonumber\\
\tilde{S}^{(V)}_{2,4}({\vec k})\! &\!=\!& -\frac12\, d z_0 V_{\vec k}
\frac{\partial z}{\partial p_0} \chi_{1}({\vec k})\,,\nonumber\\
\tilde{S}^{(V)}_{2,5}({\vec k})\! &\!=\!& -\frac12\, d V_{\vec k}\,
\chi_{0}({\vec k})\,,
\nonumber\\
\tilde{S}^{(V)}_{3,3}({\vec k})\! &\!=\!& \frac12\, \varepsilon_{V}\,.
\end{eqnarray}
where we also introduced:
\begin{equation}
\varepsilon_{V} =  V_0 n.
\end{equation}
At zero frequency the field $\psi_3$ decouples while for $V=0$ Zimmermann
\textit{et al.}'s result \cite{Zim97} is recovered.

\end{document}